\documentclass[conference, letter]{IEEEtran}

\usepackage{cite}
\usepackage{graphicx}
\usepackage{url}
\usepackage{amsmath}
\usepackage{siunitx}
\usepackage{hyperref}

\usepackage[letterpaper, left=1in, right=1in, bottom=1in, top=0.75in]{geometry}
\usepackage{tikz}
\usetikzlibrary{shapes,shapes.geometric,arrows,fit,calc,positioning,automata}
\usetikzlibrary{arrows}
\usetikzlibrary{shapes.multipart}
\usetikzlibrary{decorations.pathreplacing}
\usetikzlibrary{patterns}
\usetikzlibrary{arrows.meta}

\IEEEoverridecommandlockouts 
\newcommand\copyrighttext{%
\footnotesize \textcopyright \enspace 2018 IEEE. Personal use of this material is permitted. Permission from IEEE must be obtained for all other uses, in any current or future media, including reprinting/republishing this material for advertising or promotional purposes, creating new collective works, for resale or redistribution to servers or lists, or reuse of any copyrighted component of this work in other works. DOI: \href{https://doi.org/10.1109/PIMRC.2018.8580914}{10.1109/PIMRC.2018.8580914}
}
\newcommand\copyrightnotice{%
\begin{tikzpicture}[remember picture,overlay]
\node[anchor=south] at (current page.south) {\fbox{\parbox{\dimexpr\textwidth-\fboxsep-\fboxrule\relax}{\copyrighttext}}};
\end{tikzpicture}%
}

\begin{document}
\title{Enabling Low Latency Communications \\ in Wi-Fi Networks \thanks{The work was carried out at NRU HSE and supported by the  Russian Science Foundation (agreement 18-19-00580)}}
\author{ \IEEEauthorblockN{ Dmitry Bankov\IEEEauthorrefmark{1}\IEEEauthorrefmark{2}, Evgeny Khorov\IEEEauthorrefmark{1}\IEEEauthorrefmark{2}, Andrey Lyakhov\IEEEauthorrefmark{1}, Mark Sandal\IEEEauthorrefmark{3}}
\IEEEauthorblockA{\IEEEauthorrefmark{1}IITP RAS, Moscow, Russia, \ \ 
        \IEEEauthorrefmark{2}NRU HSE, Moscow, Russia, \ \ 
        \IEEEauthorrefmark{3}\'Ecole Polytechnique, France\\
Email: \{bankov, khorov\}@iitp.ru, mark.sandal@gmail.com}
}
\maketitle
\copyrightnotice

\begin{abstract}
	Ultra Reliable Low Latency Communications (URLLC) is an important challenge for the next generation wireless networks, which poses very strict requirements to the delay and packet loss ratio. Satisfaction is hardly possible without introducing additional functionality to the existing communication technologies. In the paper, we propose and study an approach to enable URLLC in Wi-Fi networks by exploiting an additional radio similar to that of IEEE 802.11ba. With extensive simulation, we show that our approach allows decreasing the delay by orders of magnitude, while the throughput of non-URLLC devices is reduced insignificantly.
\end{abstract}

\section{Introduction}
Ultra-Reliable Low Latency Communications (URLLC) is an integral part of the upcoming 5G wireless networks which aims at providing data transmission with very high reliability (probability of packet loss less than $10^{-5}$) and low delay (less than \SI{1}{\ms}) \cite{5G_Vision}.
Such Quality of Service requirements needs the redesign of radio access technologies. 
A good example of changes induced by URLLC is the development of the New Radio air interface for 5G cellular systems. It significantly reduces the time granularity of transmissions, since the current LTE scheduling granularity of \SI{1}{\ms} (and even greater delay between the request for channel resources and the grant) is too high to satisfy the URLLC delay requirement.

Apart from cellular networks, URLLC is also very important for Wi-Fi networks in numerous scenarios related to industrial applications, gaming, remote control, etc. Presented in Nov. 2017 at the IEEE 802 Plenary Session, the concept of Wireless Time Sensitive Networking \cite{wtsn} has obtained over 85\% affirmative votes, which reflects a great interest of Wi-Fi vendors to this topic. 
Wi-Fi does not suffer from the time granularity problem relevant to LTE, but the provision of the required delay and reliability is still a challenge, since Wi-Fi stations (STAs) use random channel access, and frame collisions cause packet losses and increase delay. 
Another important issue is that a Wi-Fi STA should wait for the channel to become idle before starting its transmission, however, transmissions of other STAs can be rather long, the upper bound being approximately \SI{5}{\ms}.
Since a Wi-Fi STA has no means to stop ongoing long frame transmission of another STA, with the current Wi-Fi it is impossible to satisfy the delay requirement of \SI{1}{\ms}.

In the paper, we refine and evaluate an approach enabling URLLC in Wi-Fi networks, proposed at the IEEE 802 Plenary Session in Nov 2017 \cite{wtsn}.
This approach is based on the usage of a secondary radio, similar to the 802.11ba one \cite{802.11ba}.
The rest of the paper is organized as follows.
In Section \ref{scheme}, we describes the proposed scheme, while in Section \ref{sec:results}, we evaluate it with simulation.
Section \ref{sec:conclusion} concludes the paper and shows future directions.

\section{Proposed Scheme} 
\label{scheme}
To access the channel, Wi-Fi devices use Enhanced Distributed Channel Access (EDCA), which is a kind of Carrier Sense Multiple Access with Collision Avoidance (CSMA/CA) with various access categories. Briefly speaking, to transmit a packet, a STA selects a random backoff value from some interval. This backoff is counted down when the channel is idle and it is frozen for the time the channel is busy plus some interframe space called AIFS. When the backoff value reaches 0, the STA transmits a frame. $SIFS$ after the transmission, the STA expects to receive an acknowledgment (ACK). If not, it considers the frame as lost and repeats transmission. 
EDCA distinguishes between four access categories (ACs): voice, video, best-effort, and background.
Each AC is assigned to a separate queue and a separate backoff function with its own contention parameters. 
By varying these parameters, some ACs are prioritized over others.

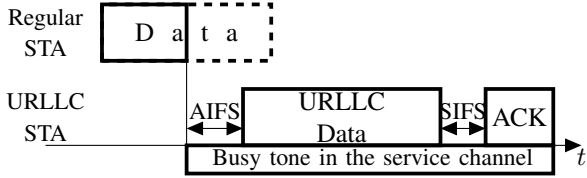
\begin{figure}[tb]
	\centering
	\begin{tikzpicture}[scale=0.75]
	\node [text width=1.8 cm, align=center] at (0,  3.0) {\small{Regular\\ STA}};
	\node [text width=1.2 cm, align=center] at (0,  1.5) {\small{URLLC STA}};
	\draw [arrows={-triangle 45}] (0,1) -- (9.5,1);
	\node at (9.5,  0.8) {$t$};
	\draw [line width=0.5mm, dashed] (1.0, 2.5) rectangle (4, 3.5);
	\draw [line width=0.5mm] (1.0, 2.5) rectangle (2.5, 3.5);
	\node [text width=1.5cm, align=center] at (2.5,  3.0) {D \ a \ t \ a \ };
	\draw [line width=0.5mm] (2.5, 0.5) rectangle (9.0, 1);
	\draw [line width=0.5mm] (3.5, 1) rectangle (7.0, 2);
	\node [text width=1.5cm, align=center] at (5.25,  1.5) {URLLC Data};
	\draw [line width=0.5mm] (7.8, 1) rectangle (9.0, 2);
	\node [text width=1.5cm, align=center] at (8.4,  1.5) {ACK};
	\draw (2.5, 1) -- (2.5, 3);
	\draw [arrows={triangle 45-triangle 45}] (2.5,1.3) -- (3.5,1.3);
	\draw [arrows={triangle 45-triangle 45}] (7.0,1.3) -- (7.8,1.3);
	\node [align=center] at (5.8,  0.75) {\small{Busy tone in the service channel}};
	\node [text width=1.0cm, align=center] at (3.0,  1.6) {\small{AIFS}};
	\node [text width=1.0cm, align=center] at (7.4,  1.6) {\small{SIFS}};
	\end{tikzpicture}
	\caption{\label{fig:access} Priority access for URLLC frames} 
	\vspace{-2em}%
\end{figure}

To enable URLLC in Wi-Fi networks, we propose to introduce an AC with the highest access priority for the URLLC traffic and to use an additional radio to provide the preemptive service.
The additional radio might be a version of the IEEE 802.11ba wake-up radio providing very robust busy-tone signaling in the narrow-band control channel, which is separate from the main data channel.
All STAs shall always listen to the control channel. As long as it is idle, the STAs may use the main channel to transmit non-URLLC frames according to the EDCA function of the corresponding AC. When some STA has a URLLC packet for transmission (below we refer to it as a URLLC STA), it starts sending a busy tone in the control channel even if some other STAs are already transmitting the busy tone.
On reception of the busy tone, all the STAs that transmit non-URLLC frames shall immediately stop ongoing transmissions and free the main channel (see Fig. \ref{fig:access}) and suspend backoff for the whole busy tone duration. Thus, we avoid collisions of URLLC frames with non-URLLC frames but do not avoid collisions inside URLLC AC. Thus, to avoid collisions of URLLC frames, apart from transmitting the busy tone, the STAs follow EDCA rules with the following exception.

If a STA starts sending the busy tone when the control channel is idle, it means that it is the only STA with URLLC frames. Thus, the STA can send the URLLC frames without waiting for backoff. Such an approach decreases delay for URLLC packets if the load is low. 

\section{Numerical Results}
\label{sec:results}

The proposed scheme reduces the delay for URLLC traffic at the price of the throughput degradation for non-URLLC one.
Moreover, it does not eliminate the contention between the URLLC STAs. For both reasons, it is important to find the maximal URLLC traffic intensity, for which the proposed approach can satisfy its QoS requirements.
For that, we consider a network with  $M$ STAs generating URLLC traffic (URLLC STAs) and $N=10$ STAs generating non-URLLC traffic (regular STAs).
Regular STAs work in the saturated mode, i.e., they always have frames for transmission.
URLLC STAs are non-saturated: after a successful transmission of a frame, a new frame arrives at a URLLC STA in a random time distributed exponentially with the mean delay of 10 ms.
We consider that all STAs can sense each other, so there are no hidden STAs.

We compare the case when all STAs are legacy with the case when all STAs support the proposed scheme. 

\begin{figure}[tb]
	\centering
	\includegraphics[width =0.9 \linewidth]{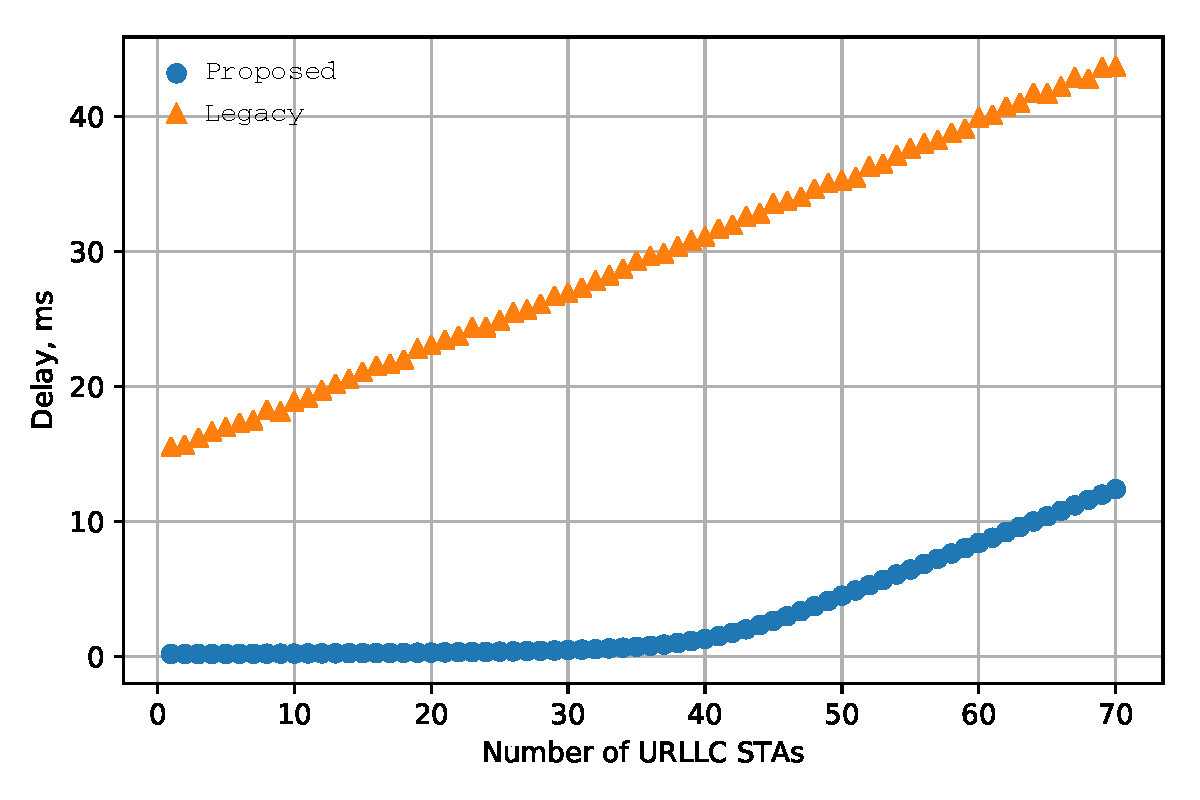}
	\includegraphics[width = 0.9\linewidth]{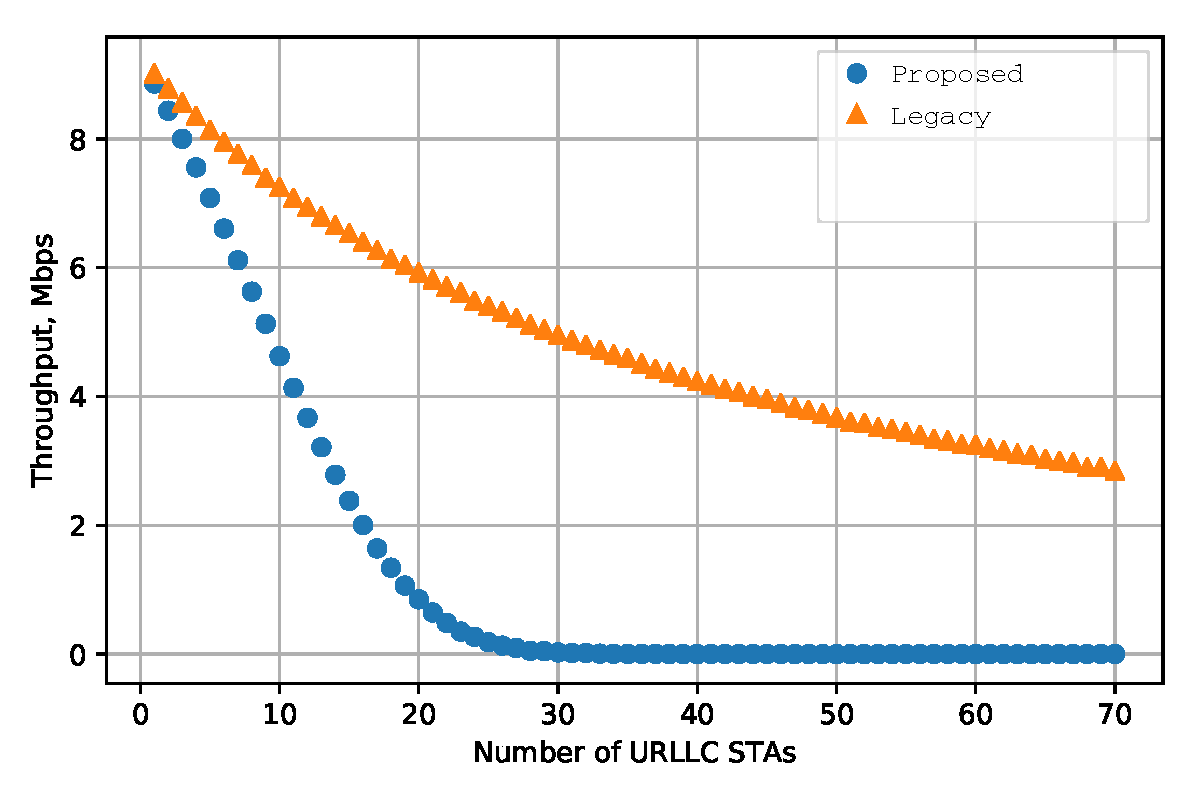}
	\caption{\label{fig:delay} Dependency of the delay for URLLC traffic and throughput of the non-URLLC one on the Number of URLLC STAs.}
	\vspace{-1.5em}%
\end{figure}

Fig.~\ref{fig:delay} shows the simulation results for the described scenario with different numbers of URLLC STAs. When the load is low, the proposed scheme manifold speeds up the transmission of URLLC frames while the throughput degradation for non-URLLC traffic is small. When the URLLC load increases, more and more non-URLLC packets are broken by URLLC transmissions. So, at 25 URLLC STAs the throughput of Regular STAs goes to 0. At this point, some channel time is still not occupied by URLLC transmissions, but the channel time fragmentation is too high. Such fragmentation does not affect URLLC STAs since a new URLLC transmission does not affect the ongoing ones. Therefore, the load for URLLC STAs can be even increased up to 35 STAs. Starting from this point, the delay significantly grows making low latency communication impossible because of very long queuing delays.  

\section{Conclusion}
\label{sec:conclusion}

In this paper, we proposed and investigate  the first-ever scheme to enable low latency communications in Wi-Fi networks. Because of paper size limitations, the results are very brief. Nevertheless, it is apparent that the proposed scheme can be implemented in future Wi-Fi networks to improve service for time-critical applications. In future works, we will study the performance of the proposed scheme in dense environments. Also, we will clarify how to service URLLC flows with different QoS requirements. Since the performance of the proposed scheme degrades at high load, we are going to improve it using OFDMA introduced in IEEE 802.11ax.

\bibliographystyle{IEEEtran}
\bibliography{biblio}

\end{document}